\PassOptionsToPackage{hypertexnames=false}{hyperref}
\documentclass{theoretics}

\usepackage{algorithm2e}

\newcommand{\cover}{\ensuremath{\mathcal{CR}_r}}

\newcommand{\para}{\paragraph}

\title{Optimal Algorithm for the Planar Two-Center Problem}

\ThCSshortnames{K.~Cho, E.~Oh,  H.~Wang, J.~Xue} 
 \ThCSshorttitle{Optimal Algorithm for the Planar Two-Center Problem}

\ThCSauthor[1]{Kyungjin Cho}{kyungjincho@postech.ac.kr}[]
\ThCSauthor[1]{Eunjin Oh}{eunjin.oh@postech.ac.kr}[]
\ThCSauthor[2]{Haitao Wang}{haitao.wang@utah.edu}[]
\ThCSauthor[3]{Jie Xue}{jiexue@nyu.edu}[]

\ThCSaffil[1]{Pohang University of Science and Technology, Korea}
\ThCSaffil[2]{University of Utah, USA}
\ThCSaffil[3]{New York University Shanghai, China}

\ThCSthanks{Invited paper from the 40th International Symposium on Computational Geometry (SoCG 2024). Kyungjin Cho and Eunjin Oh were supported by the National Research Foundation of Korea (NRF) grant funded by the Korea government (MSIT) (No.RS-2023-00209069). Haitao Wang was supported in part by NSF under Grant CCF-2300356.}

\ThCSkeywords{Two centers, disk coverage, circular hulls, \texorpdfstring{$r$}{r}-coverage} 

\ThCSyear{2024}
\ThCSarticlenum{23}
\ThCSreceived{May 1, 2024}
\ThCSrevised{Sep 30, 2024}
\ThCSaccepted{Oct 5, 2024}
\ThCSpublished{Nov 5, 2024}
\ThCSdoicreatedtrue

\begin{document}
 
\maketitle

\begin{abstract}
We study a fundamental problem in computational geometry, the \emph{planar two-center} problem.
In this problem, the input is a set $S$ of $n$ points in the plane and the goal is to find two smallest congruent disks whose union contains all points of $S$. A longstanding open problem has been to obtain an $O(n\log n)$-time algorithm for the planar two-center problem, matching the $\Omega(n\log n)$ lower bound given by Eppstein [SODA'97].
Towards this, researchers have made many efforts over decades.
The previous best algorithm, given by Wang [SoCG'20], solves the problem in $O(n\log^2 n)$ time. 
In this paper, we present an $O(n\log n)$-time (deterministic) algorithm for the planar two-center problem, which completely resolves this open problem.  
\end{abstract}

\section{Introduction} \label{sec-intro}
Given a set $S$ of $n$ points in the plane, the \emph{planar $k$-center} problem asks for $k$ smallest congruent disks in the plane whose union contains $S$.
Planar $k$-center is NP-hard if $k$ is given as part of the input.
The best known algorithm for the problem takes $n^{O(\sqrt{k})}$ time~\cite{k-center}.

In the special case where $k=1$, the problem can be solved in linear time~\cite{ref:ChazelleOn96,ref:DyerOn86,ref:MegiddoLi83,Megiddo-linear}, and this result not only holds in the plane but also in any fixed dimension. When $k=2$, the problem becomes substantially more challenging.
First of all, unlike the one-center case, the two-center problem requires $\Omega(n\log n)$ time to solve in the algebraic decision tree model, as shown by Eppstein~\cite{eppstein1997faster} by a reduction from the max-gap problem.
A longstanding open problem has been to prove a matching upper bound for the problem, i.e., obtain an $O(n \log n)$-time algorithm. Towards this, researchers have made many efforts over decades.
The first published result on the planar two-center problem appeared in 1991 by Hershberger and Suri~\cite{HERSHBERGER1991431}, who considered the decision version of the problem only and presented an $O(n^2\log n)$-time algorithm. This algorithm was subsequently improved to $O(n^2)$ time by Hershberger~\cite{ref:HershbergerA93}. Using the decision algorithm in \cite{ref:HershbergerA93} and the parametric search technique~\cite{parametric}, 
Agarwal and Sharir~\cite{AgarwalPankajK1994Pglp} presented an $O(n^2\log^3 n)$-time algorithm for the (original) two-center problem. The running time was slightly improved in a sequence of works~\cite{Eppstein-twocenter,JK-twocenter,KS-twocenter}.
A major breakthrough was achieved by Sharir~\cite{Sharir-twocenter}, who proposed an $O(n\log^9 n)$-time algorithm.
This is the first algorithm that solves the problem in near-linear time (in fact, even the first algorithm with subquadratic runtime). Eppstein~\cite{eppstein1997faster} and Chan~\cite{Chan-twocenter} further improved this result, obtaining an $O(n\log^2 n)$-time randomized algorithm and an $O(n \log^2 n \log\log n)$-time deterministic algorithm, respectively.
Since then, no progress had been made for over two decades, until recently in SoCG~2020 Wang~\cite{wang2022planar} gave an $O(n\log^2 n)$-time deterministic algorithm\footnote{Independently, Tan and Jiang~\cite{ref:TanSi17} also claimed a simple $O(n\log^2 n)$-time algorithm, but unfortunately that algorithm was later found to be incorrect~\cite{wang2022planar}.}.

Choi and Ahn~\cite{choi2021efficient} considered a special case of the two-center problem where the two disks in an optimal solution overlap and a point in their intersection is given in the input (we refer to it as the {\em anchored case}); they gave an $O(n\log n)$-time algorithm by improving the parametric search scheme in~\cite{wang2022planar}, where an algorithm of $O(n\log n\log\log n)$ time is provided for this case. In addition, Choi and Ahn's technique also solves in $O(n\log n)$ time the convex case in which the points of $S$ are in a convex position; this again improves the $O(n\log n\log\log n)$-time result in \cite{wang2022planar}.  In spite of this progress on special cases, prior to this work, Wang's $O(n\log^2 n)$-time (deterministic) algorithm~\cite{wang2022planar} remains the best known result for the general planar two-center problem.

\para{Our result.}
In this paper, we resolve this longstanding open problem completely, by presenting an $O(n\log n)$-time deterministic algorithm for the planar two-center problem.
Remarkably, in terms of running time, this is the first improvement since the $O(n\log^2 n)$-time (randomized) algorithm of Eppstein~\cite{eppstein1997faster} in 1997, although Wang~\cite{wang2022planar} later presented a deterministic version with the same time complexity.
Our result follows from a new decision algorithm for the problem which runs in $O(n)$ time after $O(n \log n)$-time preprocessing, summarized in Theorem~\ref{thm-main} (unless otherwise stated, all time complexities in the following discussion are deterministic.)
Along with known techniques in the previous works, such a decision algorithm immediately leads to an $O(n \log n)$-time algorithm for the two-center problem, as elucidated in the proof of Corollary~\ref{coro-main}.
\begin{theorem} \label{thm-main}
Let $S$ be a set of $n$ points in the plane.
After a preprocessing step on $S$ in $O(n \log n)$ time, for any given $r \geq 0$, one can compute in $O(n)$ time 
two congruent disks of radius $r$ in the plane that together covers $S$, or decide the nonexistence of two such disks.
\end{theorem}

\begin{corollary}\label{coro-main}
The planar two-center problem can be solved deterministically in $O(n \log n)$ time.
\end{corollary}
\begin{proof}
Consider an instance $S$ of the two-center problem, which is a set of $n$ points in the plane.
Let $\mathsf{opt}$ be the radius of the two disks in an optimal solution for $S$, and $\delta$ be the smallest distance between the two disk centers in any optimal solution for $S$. 

If $\delta \leq \frac{3}{2} \cdot \mathsf{opt}$, then the algorithm of Choi and Ahn~\cite{choi2021efficient} can compute an optimal solution for $S$ in $O(n \log n)$ time (in this case it is possible to sample $O(1)$ points in the plane (not necessarily from $S$) so that at least one point is in the intersection of the two optimal disks; thus the problem reduces to the anchored case).

For the case $\delta > \frac{3}{2} \cdot \mathsf{opt}$, Eppstein~\cite{eppstein1997faster} showed that if one can decide whether $r \geq \mathsf{opt}$ for any given $r$ in $T(n)$ time after $T_0(n)$-time preprocessing on $S$, then one can compute an optimal solution for $S$ in $O(T(n) \cdot \log n + T_0(n))$ time by Cole's parametric search~\cite{ColeSl87}.
The algorithm of Theorem~\ref{thm-main} solves this decision problem with $T(n) = O(n)$ and $T_0(n) = O(n \log n)$.
Therefore, we can compute an optimal solution for $S$ in $O(n \log n)$ time.

Combining the two cases finally gives us an $O(n \log n)$-time algorithm for the planar two-center problem.
Note that since we do not know which case holds for $S$,
we run the algorithms for both cases and return the better solution.
\end{proof}

To prove Theorem~\ref{thm-main}, which is the main goal of this paper, we design an algorithm that is relatively simple.
However, the correctness analysis of the algorithm is technical, entailing a nontrivial combination of several novel and intriguing insights into the problem, along with known observations in the literature.

\para{Other related work.}
A number of variants of the two-center problem have also attracted much attention in the literature, e.g., \cite{agarwal2008efficient,AgarwalPK1998TD2P,arkin2015bichromatic,de2013kinetic,halperin20022,OhBA19,OhCA18,wang2022improved}.
For example, if the centers of the two disks are required to be in $S$, the problem is known as the \textit{discrete} two-center problem. Agarwal, Sharir, and Welzl~\cite{AgarwalPK1998TD2P} solved the problem in $O(n^{4/3}\log^5 n)$ time; the logarithmic factor in the runtime was slightly improved recently by Wang~\cite{ref:WangUn23}.
An outlier version of the problem was studied in \cite{agarwal2008efficient}. 
Other variants include those involving obstacles~\cite{halperin20022,OhBA19,OhCA18}, the bichromatic version~\cite{arkin2015bichromatic,wang2022improved}, the kinetic version~\cite{de2013kinetic}, etc.
It should be noted that although the general planar $k$-center problem is NP-hard~\cite{k-center},  Choi, Lee, and Ahn recently gave a polynomial-time algorithm for the case where the points are given in convex position~\cite{ChoiLA23}. 

\para{Outline.}
The rest of the paper is organized as follows. In Section~\ref{sec-pre}, we define some basic notions that will be used throughout the paper. Section~\ref{sec-coverage} introduces a new concept, called {\em $r$-coverage}, which serves an important role in our algorithm. To the best of our knowledge, we are not aware of any previous work that used this concept to tackle the two-center problem before. We prove several properties about the $r$-coverage, which may be interesting in their own right. 
Our algorithm for Theorem~\ref{thm-main} is presented in Section~\ref{sec:algo}, while its correctness is proved in Section~\ref{sec:analysis}.
Finally, in Section~\ref{sec-conclusion}, we conclude the paper and pose some open questions.

\section{Preliminaries} \label{sec-pre}

\para{Basic notions.}
For two points $p$ and $q$ in the plane, we denote by $\overline{pq}$ the line segment connecting $p$ and $q$, and denote by $|pq|$ the distance between $p$ and $q$ (throughout the paper, ``distance'' always refers to the Euclidean distance). 
For a compact region $R$ in the plane, we use $\partial R$ to denote its boundary.
For a disk $D$ in the plane, let $\mathsf{ctr}(D)$ and $\mathsf{rad}(D)$ denote the center and the radius of $D$, respectively.
For a point $p \in \partial D$ where $D$ is a disk, the \textit{antipodal point} of $p$ on $\partial D$ refers to the (unique) point of $\partial D$ that is farthest from $p$; we often use $\hat{p}$ to denote it.
For two disks $D$ and $D'$ in the plane, denote by $\mathsf{dist}(D,D')$ the distance between the two centers $\mathsf{ctr}(D)$ and $\mathsf{ctr}(D')$.
For a number $r \geq 0$, we say that a set $Q$ of points in the plane is \textit{$r$-coverable} if there exists a radius-$r$ disk that contains all points of $Q$. 

\para{Circular hulls.}
The \emph{$r$-circular hull} of a point set $Q$ in the plane with respect to a value $r>0$ is defined as the common intersection of all radius-$r$ disks that contain $Q$ if such disks exist~\cite{ref:EdelsbrunnerOn83,HERSHBERGER1991431}. We use $\alpha_r(Q)$ 
to denote the $r$-circular hull of $Q$. See Figure~\ref{fig:coverable}(a).
Note that $Q$ is $r$-coverable if and only if $\alpha_r(Q)$ exists and is not empty.

Wang \cite{wang2022planar} considered the problem of dynamically tracking the circular hull of a set $Q$ of points in the plane under \textit{monotone} insertions, that is, each inserted point is always to the right of all points in the current $Q$. The following result is obtained in \cite{wang2022planar}.
\begin{lemma}[Theorem 6.7 in \cite{wang2022planar}]\label{lem:comp_circular_hull}
Let $Q$ be a (dynamic) set of points in the plane which is initially empty.
For any fixed number $r > 0$, there exists a data structure that maintains $\alpha_r(Q)$ in $O(1)$ amortized update time under monotone insertions to $Q$.
\end{lemma}

More specifically, $\alpha_r(Q)$ is maintained in a linked list so that it can be output in time linear in its size~\cite{wang2022planar}.

\para{Solutions for the two-center problem.}
For any $r>0$, an \textit{$r$-solution} for a planar two-center instance $S$ is a set of two radius-$r$ disks in the plane whose union covers $S$; a \textit{solution} for $S$ refers to an $r$-solution with some $r \geq 0$.
We denote by $\mathsf{opt}(S)$ the minimum $r$ such that $S$ has an $r$-solution.
An $r$-solution $\{D,D'\}$ for $S$ is \textit{tight} if $\mathsf{dist}(D,D')$ is minimized (among all $r$-solutions).
An $r$-solution $\{D,D'\}$ for $S$ is \textit{$p$-anchored} for a point $p \in \mathbb{R}^2$ if $p \in D \cap D'$.
The \textit{anchored} two-center problem takes as input a set $S$ of points in the plane and a point $p \in \mathbb{R}^2$, and asks for a $p$-anchored $r$-solution for $S$ with the minimum $r$. As discussed in Section~\ref{sec-intro}, Choi and Ahn~\cite{choi2021efficient} presented an $O(n\log n)$-time algorithm for the anchored problem.

\begin{lemma}[Choi and Ahn~\cite{choi2021efficient}] \label{lem-anchored}
The \textit{anchored} planar two-center problem can be solved by a deterministic algorithm which runs in $O(n \log n)$ time.
\end{lemma}

\section{\texorpdfstring{$\boldsymbol{r}$}{r}-Coverage} \label{sec-coverage}

We introduce a new concept, called {\em $r$-coverage}, that is critical to our algorithm. To the best of our knowledge, we are not aware of any previous work that used it to tackle the two-center problem before. We prove several properties about it, which are needed for our algorithm. We believe these properties are interesting in their own right and will find applications elsewhere. 

\begin{definition}
For a point set $Q$ in the plane and a value $r> 0$, we define the \emph{$r$-coverage} of $Q$ as the union of all radius-$r$ closed disks containing $Q$.
We use $\cover(Q)$ to denote the $r$-coverage of $Q$. 
\end{definition}

 \begin{figure}
		\begin{center}
			\includegraphics[width=0.65\textwidth]{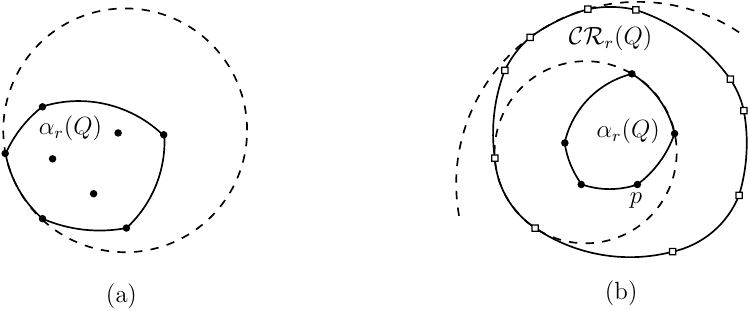}
			\caption{\label{fig:coverable}\small
				(a) Illustrating the $r$-circular hull $\alpha_r(Q)$, which is bounded by the solid arcs. The radius of the dashed circle is $r$.    
                (b) Illustrating the $r$-coverage $\cover(Q)$ (bounded by the outer cycle of solid arcs) for the set of five black points, while the inner cycle of solid arcs bounds $\alpha_r(Q)$. The larger dashed arc is of radius $2r$ while the smaller dashed circle is of radius $r$.
			}
		\end{center}
	\end{figure}

In the following, we prove several properties about $\cover(Q)$ based on an interesting relationship with $\alpha_r(Q)$. More specifically,  
we will show that $\cover(Q)$ is convex and $\partial \cover(Q)$ consists of circular arcs of radii $r$ and $2r$ alternately. Roughly speaking, these arcs correspond to the vertices and arcs of $\alpha_r(Q)$ in the following way: Each arc of radius $r$ of $\partial \cover(Q)$ is an ``antipodal arc'' of an arc of $\alpha_r(Q)$ and each arc of radius $2r$ of $\partial \cover(Q)$ has a vertex of $\alpha_r(Q)$ as its center (see Figure~\ref{fig:coverable}(b)); further, the cyclic order of the vertices and arcs of $\alpha_r(Q)$ on $\partial \alpha_r(Q)$ is consistent with the order of their corresponding arcs of $\cover(R)$ on $\partial \cover(R)$. As such, given $\alpha_r(Q)$, $\cover(Q)$ can be constructed in time linear in the number of vertices of $\alpha_r(Q)$. We formally prove these properties below. 

We assume that $Q$ can be covered by a radius-$r$ disk since otherwise $\cover(Q)$ would not exist. 
For ease of exposition, we assume that the distance of any two points of $Q$ is strictly less than $2r$ (otherwise, there is a unique radius-$r$ disk covering $Q$, which itself forms $\cover(Q)$). We start with the following lemma. 

\begin{lemma}\label{lem:boundary}
For a point $p$ on $\partial \alpha_r(Q)$, let $D$ be a radius-$r$ disk containing $Q$ with $p\in \partial D$. Then, the antipodal point of $p$ on $\partial D$ must lie on $\partial \cover(Q)$.
\end{lemma}
\begin{proof}
First of all, by the definition of $\alpha_r(Q)$, such a disk $D$ must exist for $p\in \partial \alpha_r(Q)$.
In this proof, we let $q$ be the antipodal point of $p$ on $\partial D$, and show that $q$ lies on $\partial \cover(Q)$.

For contradiction, we assume that $q$ is not on $\partial \cover(Q)$.
Then, $q$ must be in the interior of $\cover(Q)$, since $q$ lies in $\cover(Q)$ by the definition of $\cover(Q)$. Consider the ray $\rho$ from $p$ to $q$. Since $q$ is in the interior of $\cover(Q)$, $\rho$ contains a point $q'\in \cover(Q)$ such that $q$ is in the interior of $\overline{pq'}$. Since $q$ is the antipodal point of $p$ on $\partial D$ and the radius of $D$ is $r$, we have $|pq|=2r$. Hence, $|pq'|>2r$. Since $q'\in \cover(Q)$, there exists a radius-$r$ disk $D'$ containing $Q$ and $q'$. Since $D'$ contains $Q$, $D'$ contains $\alpha_r(Q)$ as well. As $p\in \alpha_r(Q)$, we obtain that $D'$ contains both $p$ and $q'$. But this is impossible as 
the radius of $D'$ is $r$ while $|pq'|>2r$. 
\end{proof}

For any point $p\in \partial \alpha_r(Q)$, define $\tau(p)$ as the set of points $q$ satisfying the conditions of Lemma~\ref{lem:boundary}, i.e., there is a radius-$r$ disk $D$ containing $Q$ and having both $p$ and $q$ on its boundary as antipodal points. By Lemma~\ref{lem:boundary}, $\tau(p)\subseteq \partial \cover(Q)$. 

Let $e$ be an arc of $\alpha_r(Q)$ with two endpoints $u$ and $v$. By slightly abusing the notation, define $\tau(e)=\cup_{p\in e}\tau(p)$. Let $D$ be the unique radius-$r$ disk whose boundary contains $e$. Note that $D$ contains $\alpha_r(Q)$ and thus contains $Q$. 
Therefore, $\tau(e)$ consists of the antipodal points of all points of $e$ on $\partial D$.
We call $\tau(e)$ the {\em antipodal arc} of $e$. Note that $\tau (e)$ and $e$ are congruent to each other (and thus they have the same length). Since we have assumed that the distance of any two points of $Q$ is less than $2r$, $\tau(e)$ does not intersect $e$. 
For later reference purposes, we have the following corollary, following directly from Lemma~\ref{lem:boundary}. 

\begin{corollary}\label{coro:antiarc}
    For an open arc $e$ of $\alpha_r(Q)$, $\tau(e)$ is an arc lying on $\partial\cover(Q)$.
\end{corollary}

 \begin{figure}
		\begin{center}
			\includegraphics[width=0.4\textwidth]{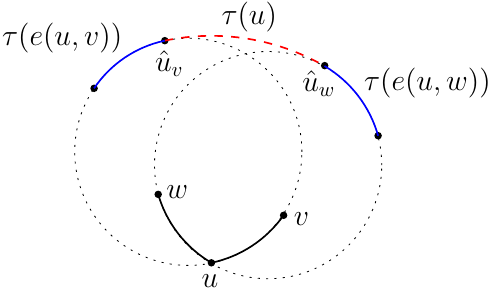}
			\caption{\label{fig:arcs}\small
	The two solid arcs connecting $v,u,w$ are $e(u,v)$ and $e(u,w)$, respectively. 			
    The red dashed arc is $\tau(u)$. The two dotted circles contain $e(u,w)$ and $e(u,v)$, respectively. The two blue solid arcs on the two circles are $\tau(e(u,v))$ and $\tau(e(u,w))$, respectively, and they share endpoints $\hat{u}_v$ and $\hat{u}_w$ with $\tau(u)$. 
			}
		\end{center}
	\end{figure}

Let $u$ be a vertex of $\alpha_r(Q)$. We now give a characterization of $\tau(u)$. 
Let $v$ and $w$ be the counterclockwise and clockwise neighboring vertices of $u$ on $\alpha_r(Q)$, respectively. See Figure~\ref{fig:arcs}. Let $e(u,v)$ be the open arc of $\alpha_r(Q)$ connecting $u$ and $v$. Let $\hat{u}_v$ be the antipodal point of $u$ on the circle containing $e(u,v)$. Define $e(u,w)$ and $\hat{u}_w$ similarly. Observe that $\tau(u)$ is the arc from $\hat{u}_v$ clockwise to $\hat{u}_w$ on the radius-$(2r)$ circle centered at $u$. Also observe that the clockwise endpoint of $\tau(u)$, which is $\hat{u}_w$, is actually the counterclockwise endpoint of $\tau(e(u,w))$; similarly, the counterclockwise endpoint of $\tau(u)$, which is $\hat{u}_v$, is the clockwise endpoint of $\tau(e(u,v))$. Hence, if we consider the arcs $e$ and vertices $u$ of $\alpha_r(Q)$ in cyclic order on $\partial \alpha_r(Q)$, their corresponding arcs $\tau(\cdot)$ form a closed cycle, denoted by $\Lambda_r(Q)$. By Lemma~\ref{lem:boundary}, $\Lambda_r(Q)\subseteq \partial \cover(Q)$. In the following, we show that $\partial \cover(Q)\subseteq \Lambda_r(Q)$ (and thus $\partial \cover(Q)= \Lambda_r(Q)$). This is an almost immediate consequence of the following lemma.

	\begin{lemma}\label{lem:antipodal}
	   Let $q$ be a point on $\partial \cover(Q)$. Then there is a unique disk $D$ of radius $r$ covering $Q$ with $q\in\partial D$. Moreover, the antipodal point $p$ of $q$ on $\partial D$ lies on $\partial \alpha_r(Q)$.
	\end{lemma}
		\begin{proof}
	    Consider a disk $D$ of radius $r$  covering $Q$ and $q\in\partial D$. Such a disk always exists by the definition
	    of $\cover(Q)$. Since $D$ covers $Q$, we have $\alpha_r(Q)\subseteq D$.
     
     We first show that the
	    antipodal point $p$ of $q$ on $\partial D$ lies on $\partial \alpha_r(Q)$. Assume to the contrary that this is not the case. Then, 
	    $p$ must lie outside of $\alpha_r(Q)$ since $\alpha_r(Q) \subseteq D$. Thus
	    there is another disk $D'$ of radius $r$ containing $Q$ but not containing $p$. 
		As such, $Q\subseteq D\cap D'$ and $p\notin D\cap D'$.
		See Figure~\ref{fig:boundary}.
  Since $Q\subseteq D'$, we have $D'\subseteq \cover(Q)$. 
  As $q\in \partial \cover(Q)$, $q$ cannot be in the interior of $D'$. 
  Further, since $p$ and $q$ are antipodal points on $\partial D$, it is possible to slightly change $D$ to a new position to obtain another radius-$r$ disk $D''$ such that $D''$ contains $D\cap D'$ (and thus contains $Q$) and also contains $q$ in its interior (e.g., first slightly rotate $D$ around $q$ towards the interior of $D'$, and then slightly translate the disk along the direction from the center toward $q$). Since $Q\subseteq D''$, it holds that $D''\subseteq \cover(Q)$. As $q$ is in the interior of $D''$, $q$ is also in the interior of $\cover(Q)$. But this leads to a contradiction since $q$ lies on $\partial \cover(Q)$.
             \begin{figure}
		\begin{center}
			\includegraphics[width=0.3\textwidth]{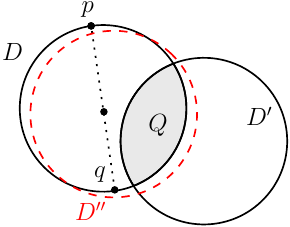}
			\caption{\label{fig:boundary}\small		 
				Illustrating $D$, $D'$, and $D''$ (the red dashed circle). $Q$ is contained in the gray region, which is $D\cap D'$. 
			}
		\end{center}
	\end{figure}
		
		We proceed to show that such a disk $D$ is unique. Assume there is another radius-$r$ disk $D'$ with $Q\subseteq D'$ and $q\in\partial D'$. Then the antipodal point $p$ of $q$ on $\partial D$ lies outside of $D'$. On the other hand, since $p$ lies on $\partial \alpha_r(Q)$ (which has been proved above), $p$ must be contained in $D'$ because $D'$ contains $Q$ (and thus contains $\alpha_r(Q)$). This leads to a contradiction. 
	\end{proof}

By Lemma~\ref{lem:antipodal}, for any point $q\in \partial\cover(Q)$, $q$ must be in $\tau(p)$ for a point $p\in \partial \alpha_r(Q)$. As such, $q\in \Lambda_r(Q)$. This also proves that 
 $\partial\cover(Q)\subseteq \Lambda_r(Q)$. We thus obtain $\partial \cover(Q)= \Lambda_r(Q)$. 
We summarize these in the following corollary (note that the arc $\tau(e)$ for each arc $e$ and $\tau(u)$ for each vertex $u$ of $\alpha_r(Q)$ can be computed in $O(1)$ time). 

\begin{corollary}\label{coro:computecover}
We have the following relations between $\cover(Q)$ and $\alpha_r(Q)$.
\begin{enumerate}[label={(\arabic*)}]
    \item $\partial \cover(Q)$ is the union of the arcs $\tau(u)$ and $\tau(e)$ for all vertices $u$ and arcs $e$ of $\alpha_r(Q)$ following their cyclical order on $\partial \alpha_r(Q)$.
    \item 
     Given $\alpha_r(Q)$, $\cover(Q)$ can be computed in time linear in the number of vertices of $\alpha_r(Q)$.
\end{enumerate}
\end{corollary}

\begin{lemma}\label{lem:coverconvex}
$\cover(Q)$ is convex.    
\end{lemma}
\begin{proof}
One way to prove this is to examine the definition of $\Lambda_r(Q)$ by observing that the three arcs $\tau(u)$, $\tau(e(u,v))$, and $\tau(e(u,w))$ always form a convex chain, for any consecutive vertices $v,u,w$ of $\alpha_r(Q)$; see Figure~\ref{fig:arcs}.
Here we give a different proof, which provides an alternative view of $\cover(Q)$.

For each $q \in Q$, let $D_q$ be the radius-$r$ disk centered at $q$.
Note that a radius-$r$ disk contains $Q$ if and only if its center lies in the common intersection $\bigcap_{q \in Q} D_q$.
Therefore, $\cover(Q)$ is the union of all radius-$r$ disks with centers in $\bigcap_{q \in Q} D_q$.
Equivalently, $\cover(Q)$ is the Minkowski sum of $\bigcap_{q \in Q} D_q$ and the radius-$r$ disk centered at the origin. Note that $\bigcap_{q \in Q} D_q$ is convex. 
As the Minkowski sum of two convex bodies is also convex, we obtain that $\cover(Q)$ must be convex.
\end{proof}
	
\section{Algorithm of Theorem~\ref{thm-main}}\label{sec:algo}

In this section, we prove Theorem~\ref{thm-main}. As discussed in Section~\ref{sec-intro}, compared to most of the previous work on the planar two-center problem, our algorithm is relatively simple, at least conceptually. However, proving its correctness is highly intricate and we devote the entire Section~\ref{sec:analysis} to it. 

Let $S$ be a set of $n$ points in the plane.
Without loss of generality, we assume that the minimum enclosing disk of $S$ is the unit disk with equation $x^2 + y^2 \leq 1$.
Choose a sufficiently large constant integer $c$ (say $c = 100$).
Set $\mathbb{Z}_{/c} = \{z/c: z \in \mathbb{Z}\}$ and $\theta = \frac{2 \pi}{c}$.
Define $\varGamma = \{(\cos i \theta,\sin i \theta): i \in [c]\}$, which is a set of $c$ unit vectors that evenly partition any disk centered at the origin.

\para{Preprocessing.}
At the outset, we compute a farthest pair $(a,b)$ of points in $S$. Let $o$ be the midpoint of the segment $\overline{ab}$.
Define
\begin{equation*}
    A = \{o\} \cup \{(x,y) \in \mathbb{Z}_{/c}^2: x^2+y^2 \leq {4}\}.
\end{equation*}
Clearly, $|A| = O(1)$.
In other words, $A$ consists of $o$ and 
all the possible $O(1)$ points of $\mathbb{Z}_{/c}^2$ in the radius-$2$ disk $x^2+y^2 \leq {4}$.

For each point $p \in A$, we compute an optimal $p$-anchored solution for $S$ by Lemma~\ref{lem-anchored}.
Among all these solutions, we take the one with smallest radius, and denote it by $\{D_0,{D_0}'\}$.
Next, for each unit vector $\vec{\gamma} \in \varGamma$, we sort the points in $S$ along the direction $\vec{\gamma}$;  let $S_{\vec{\gamma}}$ denote the corresponding sorted sequence.
This completes the preprocessing of our algorithm. 

\para{Decision procedure.}
Given a number $r> 0$, our goal is to determine whether there exist two radius-$r$ disks whose union covers $S$, and if yes, return two such disks.  
For a sequence $L$ of points in the plane, we define the \textit{maximal $r$-coverable prefix} of $L$, denoted by $\varPhi_r(L)$, as the longest prefix of $L$ that is $r$-coverable. 
For a region $R$ in the plane, we let $L\cap R$ denote the subsequence (not necessarily contiguous) of $L$ consisting of the points inside $R$. 

Algorithm~\ref{alg-decision} gives the entire algorithm.
We begin with checking whether $r \geq \mathsf{rad}(D_0)$ (note that $\mathsf{rad}(D_0)=\mathsf{rad}({D_0}')$). 
If yes, we simply return $\{D_0,{D_0}'\}$ (Line~\ref{ln-compare}).
Otherwise, we consider the vectors in $\varGamma$ one by one (the for-loop in Lines~\ref{ln-forbegin}-\ref{ln-forend}).
For each $\vec{\gamma} \in \varGamma$, we do the following.
First, we compute $X$, the maximal $r$-coverable prefix of $S_{\vec{\gamma}}$, and construct its $r$-coverage $\mathcal{CR}_r(X)$.
Then, we compute $Y$, the maximal $r$-coverable prefix of $S_{\vec{\gamma}} \cap \mathcal{CR}_r(X)$, and construct $\mathcal{CR}_r(Y)$.
Finally, we compute $Z = S_{\vec{\gamma}} \cap \mathcal{CR}_r(Y)$.
Observe that $X\subseteq Y\subseteq Z$.
Now we partition $S$ into two subsets, $Z$ and $S \setminus Z$ (here we view $Z$ as a set instead of a sequence).
We simply check whether $Z$ and $S \setminus Z$ are both $r$-coverable.
If so, we find a radius-$r$ disk $E$ (resp., $E'$) that covers $Z$ (resp., $S \setminus Z$) and return $\{E,E'\}$ as the solution.
If no solution is found after all vectors in $\varGamma$ are considered, then we return NO (Line~\ref{ln-no}).
Figure~\ref{fig:example} illustrates the algorithm with a simple example.


\para{Time analysis.}
We first show that the preprocessing can be implemented in $O(n \log n)$ time.
Finding a farthest pair $(a,b)$ of $S$ takes $O(n \log n)$ time.
For each $p \in A$, computing the optimal $p$-anchored solution can be done in $O(n \log n)$ time by Lemma~\ref{lem-anchored}.
For each $\vec{\gamma} \in \varGamma$, computing the sorted sequence $S_{\vec{\gamma}}$ takes $O(n \log n)$ time.
Since $A$ and $\varGamma$ are both of constant size, the overall preprocessing time is $O(n \log n)$.
Then we consider the time cost of the decision procedure.

\newpage 

 \begin{algorithm}[ht]
 	\caption{$\textsc{Decide}(S,r,\{D_0 ,{D_0}'   \},\{S_{\vec{\gamma}}: \vec{\gamma} \in \varGamma\})$}\label{alg-decision}
 		\If{$r \geq \mathsf{rad}(D_0)$}{  \Return{$\{D_0,{D_0}'\}$}\label{ln-compare}}
 			\For{every $\vec{\gamma} \in \varGamma$\label{ln-forbegin}}{
 	 $X \leftarrow \varPhi_r(S_{\vec{\gamma}})$\label{ln-x}\;
 		$Y \leftarrow \varPhi_r(S_{\vec{\gamma}} \cap \mathcal{CR}_r(X))$\label{ln-y}\;
 	$Z \leftarrow S_{\vec{\gamma}} \cap \mathcal{CR}_r(Y)$\label{ln-z}\; 
 		\If{there exist two radius-$r$ disks $E \supseteq Z$ and $E' \supseteq S \setminus Z$\label{ln-twodisks}}{
 		\Return{$\{E,E'\}$}\label{ln-forend}
 	}
 		}
 	\Return{NO}\label{ln-no}
 \end{algorithm}

 \begin{cfigure}[ht]
		\begin{center}
			\includegraphics[width=0.9\textwidth]{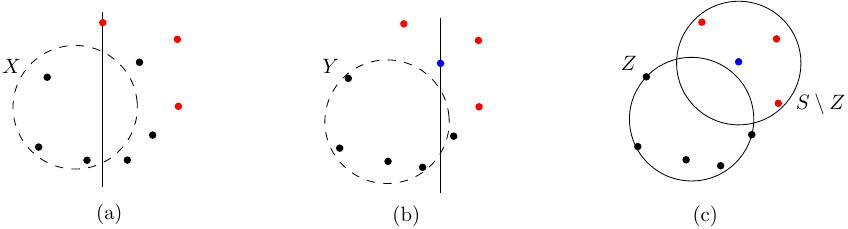}
			\caption{\label{fig:example}\small
	Illustration of Algorithm~\ref{alg-decision}. 
        (a) The red points are not in $\cover(X)$: $X$ consists of the three points inside the dashed circle.
        (b) The blue point is not in $\cover(Y)$: $Y$ consists of the four points inside the dashed circle.
        (c) The points of $Z$ and $S\setminus Z$ are covered by two radius-$r$ disks, respectively.
        Observe that $X\subseteq Y\subseteq Z$.
			}
		\end{center}
	\end{cfigure}

\begin{lemma}
Lines~\ref{ln-x}-\ref{ln-z} of Algorithm~\ref{alg-decision} can be implemented in $O(n)$ time.
\end{lemma}
\begin{proof}
For Line~\ref{ln-x}, computing $\varPhi_r(S_{\vec{\gamma}})$ can be done in linear time by Lemma~\ref{lem:comp_circular_hull} since points of $S_{\vec{\gamma}}$ are already sorted along the direction $\vec{\gamma}$. Specifically, we begin with $Q = \emptyset$ and insert points of $S_{\vec{\gamma}}$ to $Q$ one by one.
We can maintain the circular hull $\alpha_r(Q)$ in $O(1)$ amortized time using the data structure of Lemma~\ref{lem:comp_circular_hull}, as the insertions to $Q$ are monotone.
Note that $Q$ is $r$-coverable if and only if $\alpha_r(Q)$ exists.
We keep inserting points to $Q$ until $\alpha_r(Q)$ does not exist.
At this point, we obtain the maximal $r$-coverable prefix of $S_{\vec{\gamma}}$.

For Line~\ref{ln-y}, we first compute the $r$-coverage $\cover(X)$. 
Recall that we can compute $\alpha_r(X)$ in linear time using Lemma~\ref{lem:comp_circular_hull} as discussed above.
After that, $\cover(X)$ can be obtained from $\alpha_r(X)$ in linear time by Corollary~\ref{coro:computecover}. Next, we need to compute $S_{\vec{\gamma}} \cap \mathcal{CR}_r(X)$. To this end, since the $r$-coverage $\cover(X)$ is convex by Lemma~\ref{lem:coverconvex} and $S_{\vec{\gamma}}$ is sorted along $\vec{\gamma}$, computing $S_{\vec{\gamma}} \cap \mathcal{CR}_r(X)$ can be easily done in linear time, e.g., by sweeping a line perpendicular to $\vec{\gamma}$. With $S_{\vec{\gamma}} \cap \mathcal{CR}_r(X)$ available, computing $\varPhi_r(S_{\vec{\gamma}} \cap \mathcal{CR}_r(X))$ can again be done in linear time by Lemma~\ref{lem:comp_circular_hull}. 

Similarly to the above for Line~\ref{ln-y}, Line~\ref{ln-z} can also be implemented in linear time. 
\end{proof}

\noindent
The above observation implies that each iteration of the for-loop in Algorithm~\ref{alg-decision} takes $O(n)$ time, since Lines~\ref{ln-twodisks}-\ref{ln-forend} can be done in $O(n)$ time using the linear-time algorithm for the planar one-center problem~\cite{ref:ChazelleOn96,ref:DyerOn86,ref:MegiddoLi83,Megiddo-linear}, or alternatively, using Lemma~\ref{lem:comp_circular_hull}.
As $|\varGamma| = O(1)$, the for-loop has $O(1)$ iterations. As such, the total runtime of Algorithm~\ref{alg-decision} is $O(n)$.	

\section{Correctness of the algorithm}\label{sec:analysis}
In this section, we show the correctness of Algorithm~\ref{alg-decision}, which is the last missing piece for proving Theorem~\ref{thm-main}.
Before giving the analysis in Section~\ref{sec:proof}, we first prove two geometric lemmas in Section~\ref{sec:twolemmas} that will be needed in our analysis.

\subsection{Geometric lemmas}
\label{sec:twolemmas}

Let $S$ be a set of points in the plane. Recall the definition of a tight $r$-solution for $S$ in Section~\ref{sec-pre}. The following lemma describes a property of a tight $r$-solution (similar observations were made previously, e.g., \cite{Sharir-twocenter}). 

\begin{figure}[t]
		\begin{center}
			\includegraphics[width=0.45\textwidth]{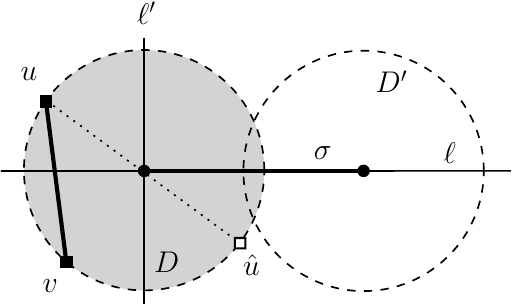}
			\caption{\label{fig:tight}\small
   Illustration of the proof of Lemma~\ref{lem-equiv}.
			}
		\end{center}
\end{figure}

\begin{lemma} \label{lem-equiv}
Let $\{D,D'\}$ be a tight $r$-solution for $S$, and $\ell$ (resp., $\sigma$) be the line (resp., open segment) connecting the two disk centers $\mathsf{ctr}(D)$ and $\mathsf{ctr}(D')$.
Then, there exist two (possibly the same) points $u,v \in (S \cap \partial D) \backslash D'$ such that 
$\overline{uv} \cap (\ell \backslash \sigma) \neq \emptyset$; see Figure~\ref{fig:tight}.
Similarly, there exist two (possibly the same) points $u',v'\in (S \cap \partial D') \backslash D$ 
such that $\overline{u'v'} \cap (\ell \backslash \sigma) \neq \emptyset$.
\end{lemma}
\begin{proof}
It suffices to find $u,v \in (S \cap \partial D) \backslash D'$ satisfying the desired condition --- the existence of $u',v'\in (S \cap \partial D') \backslash D$ can be shown likewise. Without loss of generality, we assume that $\ell$ is horizontal and $\mathsf{ctr}(D)$ is left of  $\mathsf{ctr}(D')$.

Let $\ell'$ be the line perpendicular to $\ell$ and passing through $\mathsf{ctr}(D)$. See Figure~\ref{fig:tight}. 
We define $u$ as a point in $(S \cap \partial D) \backslash D'$ on the side of $\ell'$ not containing $\sigma$, i.e., the left side of $\ell'$ by our assumption. 
Such a point always exists since $\{D, D'\}$ is a tight $r$-solution of $S$;
otherwise, we could move $D$ towards $D'$ so that $\{D,D'\}$ is still an $r$-solution while the distance between $\mathsf{ctr}(D)$ and $\mathsf{ctr}(D')$ decreases, a contradiction.
If there are multiple such points, we choose a leftmost one. 
Next, we choose $v$ as follows. 
If $u$ lies on $\ell$, we set $v=u$. In this case, $\overline{uv}$, which refers to a single point, intersects $\ell\setminus\sigma$ at the point $v$, and thus the lemma trivially holds. Below we assume $u\not\in \ell$. 

Without loss of generality, we assume that $u$ is above $\ell$. See Figure~\ref{fig:tight}. Let $\hat{u}$ be the antipodal point of $u$ on $\partial D$. 
We set $v$ as a point in $(S \cap \partial D) \backslash D'$ on the side of $\overline{u\hat{u}}$ not containing $\sigma$ and $v\not\in \overline{u\hat{u}}$. 
Such a point always exists; otherwise, we could slightly rotate $D$ counterclockwise around $u$ so that $\{D,D'\}$ is still an $r$-solution and the distance between $\mathsf{ctr}(D)$ and $\mathsf{ctr}(D')$ decreases, which contradicts the tightness of the original $\{D, D'\}$. Note that $v\neq u$ since $v\not\in \overline{u\hat{u}}$.

The above finds two different points $u$ and $v$ in $(S \cap \partial D) \backslash D'$. Now we prove 
$\overline{uv} \cap (\ell \backslash \sigma) \neq \emptyset$.
Note that $\overline{uv} \cap \sigma = \emptyset$, 
due to our choice of $v$ (recall that $v$ is on $\partial D$ that lies on the side of $\overline{u\hat{u}}$ not containing $\sigma$ and $v\not\in \overline{u\hat{u}}$).
Hence, it suffices to prove $\overline{uv} \cap \ell \neq \emptyset$.
Assume to the contrary that $\overline{uv} \cap \ell = \emptyset$. Then, $u$ and $v$ lie on the same side of $\ell$. 
As $v$ is a point on $\partial D$ that lies on the side of $\overline{u\hat{u}}$ not containing $\sigma$, $v$ must be strictly to the left of $u$. But this contradicts our choice of $u$, i.e., $u$ is a leftmost point among all points of $(S \cap \partial D) \backslash D'$ to the left of $\ell'$.
This proves $\overline{uv} \cap \ell \neq \emptyset$.
\end{proof}

The following lemma proves a property regarding two radius-$r$ disks.

\begin{lemma} \label{lem-key2}
    Let $\{D,D'\}$ be a set of two radius-$r$ disks, $u,v \in \partial D\setminus D'$ be two points whose antipodal points on $\partial D$ are not in $D'$, and $\ell$ (resp., $\sigma$) be the line (resp., open segment) connecting $\mathsf{ctr}(D)$ and $\mathsf{ctr}(D')$.
    If $\overline{uv} \cap (\ell \backslash \sigma) \neq \emptyset$, 
    then $\mathcal{CR}_r(\{u,v\}) \cap D' = D \cap D'$.
\end{lemma}
\begin{proof}
Note first that $D \subseteq \mathcal{CR}_r(\{u,v\})$ since $\{u,v\}\subseteq D$ and the radius of $D$ is $r$. 
Hence, $D \cap D' \subseteq \mathcal{CR}_r(\{u,v\}) \cap D'$.
To prove the lemma, it suffices to show $\mathcal{CR}_r(\{u,v\}) \cap D' \subseteq D \cap D'$.

\begin{figure}[t]
    \centering
    \includegraphics[height=4.5cm]{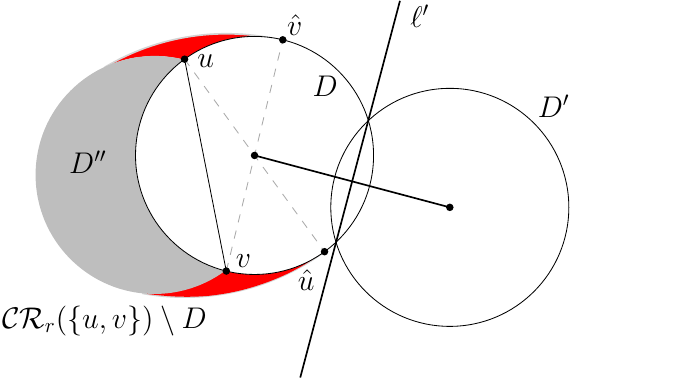}
    \caption{Illustration of the proof of Lemma~\ref{lem-key2}. The gray region is $D''\setminus D$. The two red regions are $\mathcal{CR}_r(\{u,v\})$ excluding $D\cup D''$. $\cover(\{u,v\})\setminus D$ consists of the gray region and the two red regions.}
    \label{fig-fourpts}
\end{figure}

Let $\hat{u}$ and $\hat{v}$ be the antipodal points of $u$ and $v$ on $\partial D$, respectively. By Corollary~\ref{coro:computecover}, $\cover(\{u,v\})\setminus D$ is a subset of the union of the following parts: (1) a disk $D''$, not $D$, of radius $r$ with $u$ and $v$ on its boundary, and (2) two circular sectors centered at $u$ and $v$ of radius $2r$ and tangent to $D$ and $D''$ at their antipodal points of $u$ and $v$, respectively (in Figure~\ref{fig-fourpts}, the red regions are the two circular sectors excluding the points in $D$). 
Let $\ell'$ be the bisector of the two disk centers $\mathsf{ctr}(D)$ and $\mathsf{ctr}(D')$.
Note that $u,v,\hat u,$ and $\hat v$ lie on $\partial D$ but not in $D'$ by the assumption in the lemma statement.
By definition, the union of the above parts (1) and (2) completely lies in the side of $\ell'$ containing $\mathsf{ctr}(D)$. Therefore,
$\cover(\{u,v\})\setminus D$ also completely lies in the side of $\ell'$ containing $\mathsf{ctr}(D)$. Hence, no point of $\cover(\{u,v\})\setminus D$ lies in the side of $\ell'$ containing $\mathsf{ctr}(D')$. On the other hand, every point of $D'$ in the side of $\ell'$ containing $\mathsf{ctr}(D)$ must be in $D$. It thus follows that $\mathcal{CR}_r(\{u,v\}) \cap D' \subseteq D \cap D'$.
\end{proof}

\subsection{Proving the correctness}
\label{sec:proof}
We are now in a position to prove the correctness of Algorithm~\ref{alg-decision}.
Recall our assumption that the unit disk $\{(x,y): x^2+y^2 \leq 1\}$ is the minimum enclosing disk of $S$.
In what follows, let $r > 0$ be the input value of the decision algorithm. Note that it suffices to consider the case $r<1$ since otherwise $S$ is contained in a single disk of radius $r$. 

First of all, it is clear that whenever Algorithm~\ref{alg-decision} returns a pair of disks (instead of NO), it is always an $r$-solution for $S$.
Therefore, if $r < \mathsf{opt}(S)$ (recall the definition of $\mathsf{opt}(S)$ in Section~\ref{sec-pre}), Algorithm~\ref{alg-decision} definitely returns NO.
As such, it suffices to consider the case $r \geq \mathsf{opt}(S)$. We show below that the algorithm must return an $r$-solution for $S$.

Consider a tight $r$-solution $\{D,D'\}$ for $S$.
Recall that the preprocessing step of the algorithm computes $\{D_0,{D_0}'\}$, which is the best solution among the $p$-anchored optimal solutions for all $p \in A$.
If $\{D,D'\}$ is $p$-anchored for some $p \in A$, then $r \geq \mathsf{rad}(D_0) = \mathsf{rad}({D_0}')$ and thus Algorithm~\ref{alg-decision} terminates in Line \ref{ln-compare} and returns $\{D_0,{D_0}'\}$.
In this case, the algorithm is correct.
The following lemma provides two sufficient conditions for this to happen.
\begin{lemma} \label{obs-easy}
If $\{D,D'\}$ satisfies at least one of the following two conditions, then it is $p$-anchored for some $p \in A$. 
 \begin{enumerate}[label={(\arabic*)}]
    \item $\mathsf{dist}(D,D') \leq (2-\frac{5}{c}) \cdot r$.
    \item There exist a point $u \in S \cap \partial D$ whose antipodal point on $\partial D$ is contained in $D'$ and a point $u' \in S \cap \partial D'$ whose antipodal point on $\partial D'$ is contained in $D$.
\end{enumerate}
\end{lemma}
\begin{proof}
The proof for Condition~(1) is somewhat similar to the previous work, e.g., \cite{Sharir-twocenter}; we provide some details here for completeness. 

By assumption, $S$ is contained in the unit disk $\{(x,y): x^2+y^2 \leq 1\}$ and is thus $1$-coverable. Recall that $r<1$.
Since $\mathsf{dist}(D,D') \leq (2-\frac{5}{c}) \cdot r$, $r$ is at least $1/2$. Otherwise, a single disk of radius strictly less than $1$ can cover $S$, which contradicts our assumption that the radius of the minimum enclosing disk of $S$ is $1$.
Thus, there exists a disk of radius $1/c$ contained in $D \cap D'$, and the disk in turn contains an axis-parallel square $\Box$ of side-length $1/c$.
Note that one of $D$ and $D'$ must intersect the unit disk $\{(x,y): x^2+y^2 \leq 1\}$ because $S \subseteq D \cup D'$.
As $r < 1$, we know that $\Box \subseteq \{(x,y): x^2+y^2 \leq 4\}$, and thus, $\Box \cap \mathbb{Z}_{/c}^2 \neq \emptyset$.
Therefore, $\Box$ contains some point $p \in \{(x,y) \in \mathbb{Z}_{/c}^2: x^2+y^2 \leq 4\} \subseteq A$, and thus $\{D,D'\}$ is $p$-anchored.

The proof for Condition (2) is more interesting and is new (we are not aware of any previous work that uses this condition). We assume that Condition~(1) is not satisfied, since otherwise the above already completes the proof. Thus, we have $\mathsf{dist}(D,D') > (2-\frac{5}{c}) \cdot r$.
Recall that $(a,b)$ is a farthest pair in $S$, and we included in $A$ the midpoint $o$ of the segment $\overline{ab}$.
Assuming that Condition~(2) holds, in the following we argue that $\{D,D'\}$ must be $o$-anchored, which will prove the lemma since $o\in A$.

Suppose there exist a point $u \in S \cap \partial D$ whose antipodal point on $\partial D$ is contained in $D'$ and a point $u' \in S \cap \partial D'$ whose antipodal point on $\partial D'$ is contained in $D$.
Let $\hat u$ and $\hat u'$ be the antipodal points of $u$ on $\partial D$ and $u'$ on $\partial D'$, respectively. 

For any point $p$ in the plane, we use $\vec{p}$ to denote the vector whose head is $p$ and whose tail is $\mathsf{ctr}(D)$, i.e., the center of $D$. 
Also, let $\|\vec{p}\|$ denote the magnitude of the vector $\vec p$, or equivalently, the distance between $p$ and $\mathsf{ctr}(D)$.
Note that $\|\vec{p}-\vec{q}\|=|pq|$ for any two points $p$ and $q$.
We have
\begin{align} \label{eq:dot_product}
    & \|\vec{p}+\vec{q}\|  = \sqrt{\|\vec{p}\|^2 + \|\vec{q}\|^2 + 2 \langle \vec p,\vec q \rangle}\nonumber, \ \ \ \ \ \ \ \|\vec{p}-\vec{q}\|  = \sqrt{\|\vec{p}\|^2 + \|\vec{q}\|^2 - 2 \langle \vec p,\vec q \rangle}\nonumber,\textnormal{ and}\\
    & \|\vec{p}+\vec{q}\|^2 + \|\vec{p}-\vec{q}\|^2 = 2 \|\vec{p}\|^2 + 2\|\vec{q}\|^2,
\end{align}
where $\langle \cdot,\cdot \rangle$ denotes the dot product.

We claim that neither $D$ nor $D'$ contains both $a$ and $b$. Indeed, assume to the contrary this is not the case. Then, 
$|ab|\leq 2r$. 
This implies $|uu'|\leq 2r$ since $(a,b)$ a farthest pair of $S$ and $u,u'\in S$. 
Consequently, the distance between any two points among $u, u', \hat u$, and $\hat u'$ is at most $2r$
since the antipodal points $\hat u$ and $\hat u'$ are contained in $D\cap D'$.
Note that $\mathsf{ctr}(D)$ is the midpoint of $\overline{u\hat u}$.
Analogously, $\mathsf{ctr}(D')$ is the midpoint of $\overline{u'\hat u'}$. 
As such, we have $\mathsf{dist}(D,D')\leq 3r/2$ (refer to the caption of Figure~\ref{fig:no_ab_in_one} for a detailed explanation). But this leads to a contradiction since $\mathsf{dist}(D,D') > (2-\frac{5}{c}) \cdot r>3r/2$, with $c=100$. 
Hence, neither $D$ nor $D'$ contains both $a$ and $b$.

\begin{figure}
    \begin{center}
	\includegraphics[width=0.5\textwidth]{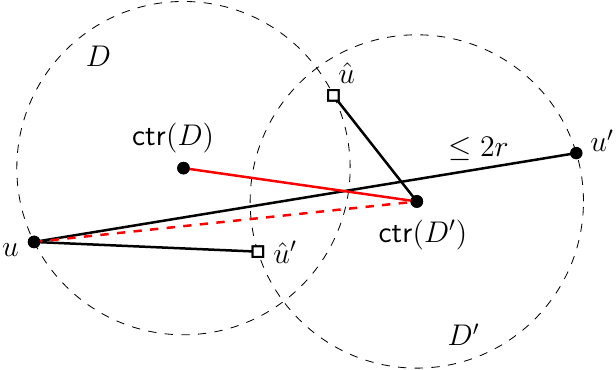}
	\caption{\label{fig:no_ab_in_one}\small
        The length of the red line segment connecting the two disk centers, which is $\mathsf{dist}(D,D')$, is at most the average of the two distances from $\mathsf{ctr}(D')$ to $u$ and $\hat u$, respectively. The distance between $u$ and $\mathsf{ctr}(D')$, which is the length of the red dashed segment, is at most $2r$ since $|u\hat u'|\leq 2r$ and $|uu'|\leq 2r$.
        Furthermore, since $\hat u\in D'$, the distance between $\hat u$ and $\mathsf{ctr}(D')$ is at most $r$.
        Therefore, $\mathsf{dist}(D,D')\leq 3r/2$. 
	}
    \end{center}
\end{figure}
Without loss of generality, we assume $b\in D$ and $a\in D'$. Recall that our goal is to show that $o\in D\cap D'$. In the following, we only prove $o\in D$ since $o\in D'$ can be proved analogously. Let $\delta=|ao|$, i.e., $\delta=|ab|/2$. 

We first show that the distance between $a$ and $\mathsf{ctr}(D)$ is at most $\sqrt{r^2+2\delta^2}$, i.e., $\|\vec a\|\leq \sqrt{r^2+2\delta^2}$. 
Indeed, since $(a,b)$ is a farthest pair of $S$ and $u,a\in S$, we have $|au|\leq |ab|=2\delta$. Because both $a$ and $\hat{u}$ are in $D'$, $|a\hat{u}|\leq 2r$. In addition, $\|\vec{ u}\|= r$ holds since $u$ lies on $\partial D$. By Equality~\eqref{eq:dot_product}, 
we have the following (note that $\|\vec{a}+\vec{u}\| = \|\vec{a}-\vec{\hat u}\| =|a\hat u| \leq 2r$): 
\begin{align*}
    2\|\vec a\|^2&=\|\vec a+\vec u\|^2+\|\vec a-\vec u\|^2-2\|\vec u\|^2 \\
    &=|a\hat u|^2+|au|^2-2r^2\\
    & \leq 4r^2 + 4\delta^2-2r^2 = 2r^2+4\delta^2.
\end{align*}

We are now in a position to prove $o\in D$. It suffices to show $\|\vec{o}\|\leq r$.
Recall that $\vec{a}-\vec{o}=\vec{o}-\vec{b}$ since $o$ is the midpoint of $\overline{ab}$. Notice that $\|\vec{b}\|\leq r$ since $b\in D$. From $\|\vec{a}\|\leq \sqrt {{r}^2+2\delta^2}$ and Equation~\eqref{eq:dot_product}, we have the following (which leads to $\|\vec o\|\leq r$):
\begin{align*}
    (r^2+2\delta^2)+r^2 \geq \|\vec{a}\|^2+ \|\vec{b}\|^2& = \|\vec{o}+(\vec{a}-\vec{o})\|^2 + \|\vec{o}+(\vec{b}-\vec{o})\|^2\\
    & = \|\vec{o}+(\vec{a}-\vec{o})\|^2 +\|\vec{o}-(\vec{a}-\vec{o})\|^2\\
    &= 2\|\vec{o}\|^2 +2\|\vec{a}-\vec{o}\|^2\\
    &= 2\|\vec{o}\|^2 +2\delta^2
\end{align*}
As such, we obtain $o\in D$. 

Similarly, we can prove $o\in D'$. Therefore, $\{D,D'\}$ is $o$-anchored. 
\end{proof}

In what follows, it suffices to focus on the case where neither condition in Lemma~\ref{obs-easy} holds.
The failure of condition~1 implies $\mathsf{dist}(D,D') > (2-\frac{5}{c}) \cdot r$.
For the failure of condition~2, every point $u \in S \cap \partial D$ has its antipodal point on $\partial D$ outside $D'$, or this holds for $D'$. Without loss of generality, we assume that this is true for $D$. 
We argue below that there exists at least one vector $\vec{\gamma} \in \varGamma$ such that when Algorithm~\ref{alg-decision} considers $\vec{\gamma}$ in the for-loop, it returns a solution in Line~\ref{ln-forend}.

\begin{figure}
    \centering
    \includegraphics[width=0.58\textwidth]{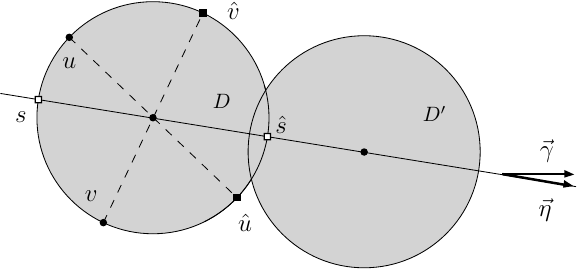}
    \caption{Illustration of $u,v, s, \hat s, \vec \gamma, \vec\eta$. For convenience, only the directions of $\vec \gamma,\vec \eta$ are presented.}
    \label{fig:notations}
\end{figure}
 
\para{Choosing $\boldsymbol{\vec{\gamma}}$.}
The choice of $\vec{\gamma}$ is as follows.
Let $\ell$ (resp., $\sigma$) be the line (resp., open segment) connecting the two centers $\mathsf{ctr}(D)$ and $\mathsf{ctr}(D')$.
For any point $p \in \partial D$, we denote by $\hat{p}$ the antipodal point of $p$ on $\partial D$.
The line $\ell$ intersects $\partial D$ at two antipodal points $s$ and $\hat{s}$, where $s$ (resp., $\hat{s}$) denotes the one farther (resp., closer) to $\mathsf{ctr}(D')$.
By Lemma~\ref{lem-equiv}, there exist $u,v \in S \cap (\partial D \setminus D')$ such that
$\overline{uv} \cap (\ell\setminus\sigma) \neq \emptyset$. Note that this implies 
$\overline{uv} \cap \ell \neq \emptyset$ and $\overline{uv}\cap \sigma=\emptyset$.
By our assumption, $u,v,\hat{u},\hat{v} \notin D'$.
Without loss of generality, we can assume that $|su| \leq |sv|$.
See Figure~\ref{fig:notations}.
The points $s$ and $\hat{s}$ partition $\partial D$ into two arcs; the \textit{$u$-arc} refers to the one where $u$ lies on. 
Since $\overline{uv} \cap \ell \neq \emptyset$, $v$ must lie on the arc other than the $u$-arc, which we call the \textit{$v$-arc}. (If $u = v = s$, 
we pick an arbitrary arc as the $u$-arc and let the other one be the $v$-arc.)
For each vector $\vec{\gamma} \in \varGamma$, we shoot a ray from $\mathsf{ctr}(D)$ with direction $\vec{\gamma}$.
These rays intersect $\partial D$ at $c$ points (which evenly partition $\partial D$ into $c$ arcs).
Among these $c$ intersection points, we pick the one on the $u$-arc that is closest to $\hat{s}$.
Then we define $\vec{\gamma} \in \varGamma$ as the vector corresponding to this intersection point.

\para{Justifying $\boldsymbol{\vec{\gamma}}$.}
We show that when considering $\vec{\gamma}$, Algorithm~\ref{alg-decision} returns a solution.
Let $\vec{\eta} \in \mathbb{S}^1$ be the unit vector with the same direction as the one originated from $\mathsf{ctr}(D)$ to $\hat{s}$.
Note that the angle between $\vec{\eta}$ and $\vec{\gamma}$ is at most $\frac{2 \pi}{c}$ by the construction of $\varGamma$ and the choice of $\vec{\gamma}$.
By rotating and reflecting the coordinate system, we can assume that $\vec{\gamma} = (1,0)$ and 
the clockwise angle from $\vec{\gamma}$ to $\vec{\eta}$ is at most $\frac{2 \pi}{c}$; see Figure~\ref{fig:notations}. 

For two points $p,q \in \mathbb{R}^2$, we write $p \prec q$ if 
the $x$-coordinate of $p$ is smaller than that of $q$.
The \emph{angle} of an arc of $\partial D$ is defined as the central angle formed by the arc at $\mathsf{ctr}(D)$.
The following observation pertains to the relative positions of certain points with respect to $u$ and $v$. 

\begin{lemma}\label{obs-3conditions}
The points $u$ and $v$ satisfy the following (see Figure~\ref{fig:notations}).
\begin{enumerate}[label={(\arabic*)}]
    \item $v \prec \hat{u}$.
    \item $u \prec p$ holds for any point $p \in D'$.
    \item $|pu|>2r$ for any point $p \in D'\setminus D$ with $p \prec \hat u$. Note that $|pu|>2r$ implies $p\not\in \mathcal{CR}_r(\{u\})$.
\end{enumerate}
\end{lemma}
\begin{proof}
We prove these three conditions in order. 

\para{Proving (1).}
To see Condition~(1), observe that the angle of the subarc within the $u$-arc between $s$ and $u$ is at most $\frac{\pi}{2}$, since otherwise $v$ would be closer to $s$ due to $\overline{uv} \cap (\ell\setminus\sigma) \neq \emptyset$, contradicting our assumption $|su|\leq |sv|$. 
As the smaller angle between $\vec{\eta}$ and $\vec{\gamma}$ is at most $\frac{2\pi}{c} < \frac{\pi}{4}$, $s\prec \hat u$ must hold.
So we have $v \prec \hat u$, since $v$ lies on the subarc of the $v$-arc between $s$ and $\hat u$.

\para{Proving (2).}
For Condition~(2), we define a function $f: \mathbb{R}^2 \rightarrow \mathbb{R}$ as $f(p) = \langle p, \vec{\gamma} \rangle - \langle \mathsf{ctr}(D'), \vec{\gamma} \rangle$, where $\langle \cdot, \cdot \rangle$ denotes the dot product.
Note that for any points $p,q$ in the plane, $p \prec q$ if and only if $f(p) < f(q)$.
It suffices to show that $f(u) < -r$, since $f(p) \geq -r$ for any $p \in D'$.
Recall that $r$ is the radius of $D$ (and of $D'$).
We then have
\begin{align*}
    f(u)
    &\leq f(s) +\frac{\pi r}2= f(\mathsf{ctr}(D)) -r\cdot \langle \vec{\eta}, \vec{\gamma} \rangle+\frac{\pi r}2\\
    &\leq -\left(3-\frac5 c\right)\cdot r \cdot \langle \vec{\eta}, \vec{\gamma} \rangle +\frac{\pi r}2 \\
    &\leq -\left(3-\frac5 c\right)\cdot \left(1-\frac{2\pi}c\right)\cdot r+\frac{\pi r}2 < -r.
\end{align*}
The first inequality holds because the distance between $s$ and $u$ is at most $\frac{\pi r}{2}$ (this is in turn because the angle of the subarc within the $u$-arc between $s$ and $u$ is at most $\frac{\pi}{2}$ as already discussed above). 
The equality in the first line holds because $\vec{\eta}$ is the unit vector directed from $s$ to $\mathsf{ctr}(D)$ and the distance between $s$ and $\mathsf{ctr}(D)$ is $r$. 
The second inequality holds since $f(\mathsf{ctr}(D))=-\mathsf{dist}(D,D') \cdot \langle \vec{\eta}, \vec{\gamma} \rangle$ and we have assumed $\mathsf{dist}(D, D') > (2 - \frac{5}{c}) \cdot r$.
The last one follows from $\langle \vec{\eta}, \vec{\gamma} \rangle \geq 1-\frac{2\pi}{c}$, as the angle between $\vec\gamma$ and $\vec\eta$ is at most $\frac{2\pi}{c}$.
This proves the second condition. 

\para{Proving (3).}
We finally prove Condition~(3). 
Let $p$ be any point in $D'\setminus D$ with $p \prec \hat u$. Our goal is to prove $|pu|>2r$.
Define $\ell'$ to be the perpendicular bisector line of $\sigma$. Let $\ell_p$ be the vertical line through $p$. 
By the way, we chose $\vec \gamma$, $u$ and the vector $\vec \gamma$ originating from $\mathsf{ctr}(D)$ lie on the same side of $\ell$.
Recall that we have rotated the coordinate system so that $\vec \gamma=(1,0)$. Hence, 
if $u$ is above $\ell$, then the direction of $\vec\eta$ is downward, i.e., $\mathsf{ctr}(D)$ is higher than $\mathsf{ctr}(D')$ (see Figure~\ref{fig-angle}(a)). 
If $u$ is below $\ell$, then $\vec\eta$ is upward and $\mathsf{ctr}(D')$ is higher than $\mathsf{ctr}(D)$ (see Figure~\ref{fig-angle}(b)).
In the following, we only consider the case where $u$ is above $\ell$, since the other case can be proved analogously. Since $\hat{u}$ is the antipodal point of $u$ on $\partial D$ and $\ell$ contains $\mathsf{ctr}(D)$, $\hat{u}$ is below $\ell$.

\begin{figure}[ht]
    \centering
    \includegraphics[height=5.0cm]{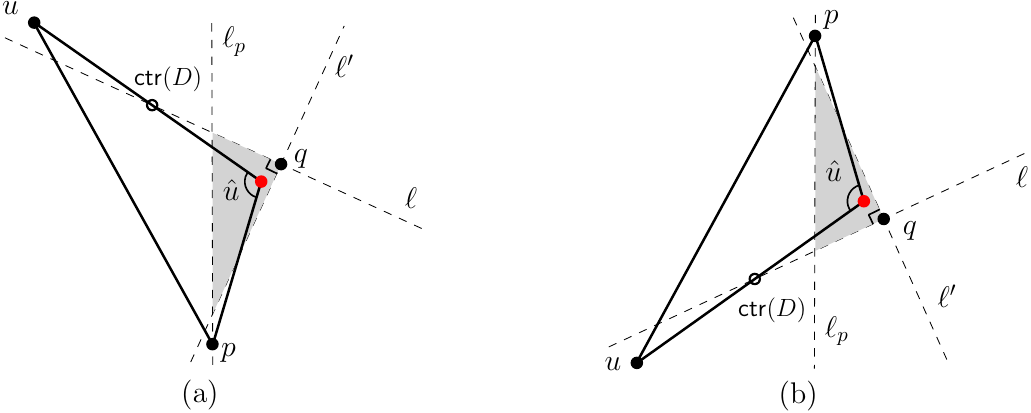}
    \caption{Illustration of the proof of Condition~(3) of Lemma~\ref{obs-3conditions}.}
    \label{fig-angle}
\end{figure} 

We claim that $\hat u$ must lie in the triangle bounded by the lines $\ell$, $\ell'$, and $\ell_p$; see Figure~\ref{fig-angle}. To see this, first of all, since $p\prec \hat{u}$, $\hat{u}$ is to the right of $\ell_p$. Also, notice that $D\setminus D'$ is to the left of $\ell'$. Since $\hat{u}\in  D$ and $\hat{u}\not\in D'$, we obtain that $\hat{u}$ is to the left of $\ell'$. 
Finally, since $\mathsf{ctr}(D)$ is higher than $\mathsf{ctr}(D')$, $\hat{u}$ is below $\ell$, $\hat{u}\in \partial D$, and $\hat{u}\not\in D'$, we obtain that the intersection $q=\ell\cap \ell'$ must be to the right of the vertical line through $\hat{u}$. Since $\hat{u}$ is right of $\ell_p$, we obtain that $q$ is to the right of $\ell_p$. The above discussions together lead to that $\hat{u}$ must be in the triangle bounded by $\ell$, $\ell'$, and $\ell_p$. 

By the definition of $\ell'$, the portion of $D'$ to the left of $\ell'$ must be inside $D$. Since $p\in D'\setminus D$, $p$ must be to the right of $\ell'$. Recall that $\hat{u}$ is to the left of $\ell'$. Since $u$ is above $\ell$ while $\hat{u}$ is below $\ell$, we obtain that 
the triangle $\Delta p\hat{u} u$ has an obtuse angle at $\hat u$; see Figure~\ref{fig-angle}. 
As a consequence, $|up|>|u\hat u|=2r$. 
\end{proof}

Consider the sets $X,Y,Z$ computed in Lines~\ref{ln-x}-\ref{ln-z} of Algorithm~\ref{alg-decision} for the vector $\vec{\gamma}$. Our goal is to prove that $Z$ is exactly equal to $S \cap D$ (when viewed as a set), which implies $S \backslash Z \subseteq D'$ because $S\subseteq D\cup D'$.
Note that as long as this is true, both $Z$ and $S \backslash Z$ are $r$-coverable, and thus Algorithm~\ref{alg-decision} will return a solution in Line~\ref{ln-forend} when considering $\vec{\gamma}$.

\begin{lemma}\label{obs-main_correctness}
Let $u$ and $v$ be as above. We have $u\in X\subseteq D$, $\{u,v\} \subseteq Y \subseteq D$, and $Z = S \cap D$.
\end{lemma}
\begin{proof}
Let $(p_1,\dots,p_n)$ be the sorted sequence $S_{\vec\gamma}$; let $j$ and $k$ be the indices with $u=p_j$ and $v=p_k$.
Recall that Algorithm~\ref{alg-decision} computes $X, Y$ and $Z$ in this order. 

\para{Proving $\boldsymbol{u\in X\subseteq D}$.}
We first prove $u\in X\subseteq D$. 
Recall that $X$ is the maximal $r$-coverable prefix of $S_{\vec \gamma}$.
According to Lemma~\ref{obs-3conditions}(2), $p_i \notin D'$ for all indices $i \leq j$, and therefore $\{p_1,\dots,p_j\} \subseteq D$. 
Hence, $\{p_1,\dots,p_j\}$ is an $r$-coverable, and thus $\{p_1,\dots, p_j\} \subseteq X$.
In particular, $u = p_j \in X$.
To show $X \subseteq D$, we distinguish two cases: $v \in X$ and $v \notin X$.
\begin{itemize}
    \item If $v \in X$, then we have $X\subseteq\mathcal{CR}_r(\{u,v\})$ since $u,v\in X$ and $X$ is $r$-coverable. By Lemma~\ref{lem-key2}, $\mathcal{CR}_r(\{u,v\}) \cap D'= D \cap D'$. As such, we obtain $X \cap D'\subseteq D \cap D'$, and consequently, $X \cap (D' \setminus D) = \emptyset$.
This further implies that $X \subseteq D$ since $X\subseteq D\cup D'$. 
\item 
If $v\notin X$, we assume $X\nsubseteq D$ for the sake of contradiction. 
Then, there exists $p_i$ such that $\{p_1,\dots,p_i\}\subseteq X$ is $r$-coverable but $p_i$ lies outside $D$ (and thus $p_i\in D' \setminus D$). Since $v\not\in X$, we have 
$p_i\prec v$. As $v\prec \hat{u}$ by Lemma~\ref{obs-3conditions}(1), 
$p_i\prec \hat u$ holds. Because $p_i\in D' \setminus D$ and $p_i\prec \hat u$, we further have 
$p_i\notin \mathcal{CR}_r(\{u\})$ by Lemma~\ref{obs-3conditions}(3).

On the other hand, since $X$ is $r$-coverable and $u\in X$, $X\subseteq \mathcal{CR}_r(\{u\})$ must hold. As $p_i\in X$, we have $p_i\in \mathcal{CR}_r(\{u\})$. We thus obtain contradiction.  
\end{itemize}

\para{Proving $\boldsymbol{\{u,v\} \subseteq Y \subseteq D}$.}
We now prove $\{u,v\} \subseteq Y \subseteq D$.
Recall that $Y$ is the maximal $r$-coverable prefix of $S_{\vec\gamma}\cap \mathcal{CR}_r(X)$.
As $X\subseteq D$, $D\subseteq \mathcal{CR}_r(X)$ holds. Since $v\in D$, we have $v\in S\cap \mathcal{CR}_r(X)$.
Furthermore, since $X$ is $r$-coverable, it follows that $X\subseteq \mathcal{CR}_r(X)$, which further implies $X\subseteq Y$. As $u\in X$, we have $u\in Y$.

To show $v \in Y$, we claim that all points of $\{p_1,\dots,p_k=v\}\cap \mathcal{CR}_r(X)$ lie in  $D$.
Assume to the contrary this is not true. Then, there is a point $p_i$ in $\{p_1,\dots,p_k=v\}\cap \mathcal{CR}_r(X)$ such that $p_i\notin D$ (and thus $p_i\in D'\setminus D$). Thus, either $p_i=v$ or $p_i\prec v$. Since $v\prec \hat{u}$ by Lemma~\ref{obs-3conditions}(1), we have $p_i\prec \hat{u}$. Since $p_i\in D' \setminus D$ and $p_i\prec \hat u$, we further have $p_i\notin \mathcal{CR}_r(\{u\})$ by Lemma~\ref{obs-3conditions}(3). On the other hand, since $u\in X$ and $X$ is $r$-coverable, $\mathcal{CR}_r(X)\subseteq \mathcal{CR}_r(\{u\})$ holds. As $p_i\in \mathcal{CR}_r(X)$, we obtain $p_i\in \mathcal{CR}_r(\{u\})$, a contradiction. 

The above claim implies that $\{p_1,\dots,p_k=v\} \cap \mathcal{CR}_r(X)$ is $r$-coverable, and thus $v$ must be in $Y$.
As $\{u,v\} \subseteq Y$ and $Y$ is $r$-coverable, it follows that $Y\subseteq \mathcal{CR}_r(\{u,v\})$.
According to Lemma~\ref{lem-key2}, $\mathcal{CR}_r(\{u,v\}) \cap D' = D \cap D'$. Therefore, $Y\cap D'\subseteq D\cap D'$, which leads to  $Y\subseteq D$ since $Y\subseteq D\cup D'$.

\para{Proving $\boldsymbol{Z=S \cap D}$.}
We finally prove $Z=S\cap D$, using the fact $\{u,v\} \subseteq Y \subseteq D$.
Recall that $Z=S \cap \mathcal{CR}_r(Y)$.
As $Y \subseteq D$ and $Y$ is $r$-coverable, we have $D \subseteq \mathcal{CR}_r(Y)$, which implies $S \cap D \subseteq Z$.
It suffices to show $Z \subseteq S \cap D$, or equivalently, $Z \subseteq D$.
Observe that $Z = (Z\cap D)\cup (Z\cap D')$, since $D \cup D'$ contains all points in $Z$.
Therefore, we only need to show $Z \cap D' \subseteq D$.
Since $\{u,v\}\subseteq Y$ and $Y$ is $r$-coverable, we have $\cover(Y)\subseteq\mathcal{CR}_r(\{u,v\})$.
By Lemma~\ref{lem-key2}, $\mathcal{CR}_r(\{u,v\}) \cap D'=D\cap D'$. 
It follows that
\begin{equation*}
    Z\cap D' \subseteq \mathcal{CR}_r(Y)\cap D'\subseteq \mathcal{CR}_r(\{u,v\})\cap D' = D\cap D' \subseteq D.
\end{equation*}
We then have $Z \cap D' \subseteq D$ and thus $Z \subseteq D$.

This completes the proof. 
\end{proof}

We have thus completed the full proof of the correctness of Algorithm 1, from which we derive Theorem 1 and Corollary 2.

\section{Conclusion and future work} \label{sec-conclusion}

This paper presents an $O(n \log n)$-time deterministic algorithm for the planar two-center problem, which improves the previous $O(n \log^2 n)$-time algorithm by Wang~\cite{wang2022planar} and matches the $\Omega(n \log n)$ lower bound for the problem given by Eppstein~\cite{eppstein1997faster}.
Our algorithm has a straightforward design while its analysis is technically involved and requires new approaches to the problem.
We believe that some of our techniques can find applications in other variants of the planar two-center problem (or more generally, other geometric coverage problems).

Next, we pose some interesting open questions for future study.
First, given that the planar two-center problem can now be solved optimally, it is natural to ask for the largest $k$ such that planar $k$-center can be solved in $O(n \log n)$ time (or near-linear time).
This question is completely open: even for $k = 3$ with the assumption that all input points are in convex positions, the best-known algorithm runs in near-quadratic time \cite{bian2018efficient}.
Furthermore, one can also study the two-center problem in higher dimensions and ask whether the problem can be solved in near-linear time in any fixed dimension (which is true for the one-center problem).
Finally, we notice that, while near-linear time algorithms for the planar two-center problem have been known for more than two decades since the work of Sharir~\cite{Sharir-twocenter}, the best known algorithms for the \textit{discrete} planar two-center problem run in $O(n^{4/3} \cdot \text{poly}(\log n))$ time~\cite{AgarwalPK1998TD2P,ref:WangUn23}, as mentioned in the introduction.
Therefore, it is interesting to ask whether the techniques in this paper can also be applied to design near-linear time algorithms for the discrete problem.


\printbibliography

\end{document}